\documentclass[condensedmatter,preprints,submit,moreauthors,dvi2pdf,10pt,a4paper]{mdpi} 
\firstpage{1} 
\articlenumber{x}
\doinum{10.3390/------}
\pubvolume{xx}
\pubyear{2017}
\copyrightyear{2017}
\history{Received: date; Accepted: date; Published: date}

\Title{Fermionic properties of two interacting bosons in a two-dimensional harmonic trap}

\Author{Pere Mujal $^{1,2,*}$\orcidA{}, Artur Polls $^{1,2}$  and Bruno Juli\'{a}-D\'{i}az $^{1,2,3}$}

\AuthorNames{Pere Mujal, Artur Polls and Bruno Juli\'{a}-D\'{i}az}

\address{%
$^{1}$ \quad Departament de F\'{i}sica Qu\`{a}ntica i Astrof\'{i}sica,
Universitat de Barcelona, Mart\'{i} i Franqu\`{e}s 1, 08028 Barcelona, Spain\\
$^{2}$ \quad Institut de Ci\`{e}ncies del Cosmos (ICCUB), Universitat 
de Barcelona, Mart\'{i} i Franqu\`{e}s 1, 08028 Barcelona, Spain\\
$^{3}$ \quad Institut de Ci\`{e}ncies Fot\`{o}niques, Parc Mediterrani 
de la Tecnologia, 08860 Barcelona, Spain}
\corres{Correspondence: peremujal@fqa.ub.edu}

\abstract{
The system of two interacting bosons in a 
two-dimensional harmonic trap is compared with the system consisting of two noninteracting fermions in the same potential. In particular, we discuss 
how the properties of the ground state of the system, e.g., the different contributions to the 
total energy, change as we vary both the strength and range of the atom--atom 
interaction. In particular, we focus on the short-range and strong interacting limit of the two-boson system and compare it to the noninteracting two-fermion system by properly symmetrizing the corresponding degenerate ground state wave functions. In that limit, we show that the density profile of the two-boson system has a tendency similar to the system of two noninteracting fermions. Similarly, the correlations induced when the interaction strength is increased result in a similar pair correlation function for both systems.
}


\begin{document}
\section{Introduction}
In one dimension, the Bose--Fermi mapping~\cite{Girardeau} theorem establishes a relation between 
the ground state energy and the wave function of strongly interacting bosons 
with those corresponding to noninteracting fermions in the same trapping 
potential. Both systems have the same energy, and the ground 
state wave function of the interacting bosonic system can be obtained by 
symmetrizing the noninteracting fermionic one by taking the absolute value. Several 
works have discussed the onset of the Tonks--Girardeau phase in 
one dimension~\cite{Girardeau2,Paredes,Kinoshita,Pupillo,Garcia-March4,Garcia-March,
Pyzh,Tempfli,Koscik,Bakr,Sherson,Busch,Deuretzbacher,Garcia-March3,Wilson,Barfknecht,Koscik2,Brouzos}.

In contrast, in two or more dimensions, the theorem does not apply because it is 
based on the  principle, valid in one dimension, that if a particle is fixed at a point, the space 
becomes completely separated in two pieces~\cite{Girardeau}. Therefore, another 
particle cannot access the whole space without encountering the one that is 
fixed, which is not the case with more dimensions. In contrast, the mechanism 
described by Girardeau whereby the particles can avoid feeling the interaction  remains a possibility. In our previous work~\cite{PRA96.043614}, 
considering two, three, and four particles~\cite{Blume} in a two dimensional harmonic trap~\cite{Busch,Doganov,Liu,Christensson,Kartavtsev,Shea,Daily,Whitehead,Farrell,Zinner}, we showed that interacting bosons avoid feeling the interaction by becoming 
correlated in such a way that the probability of two particles being at the same 
position~vanishes~\cite{Romanovsky}. 

In the present paper, we compare the numerical calculations for the ground state 
of two interacting bosons in two dimensions with a short-range interaction with the properties 
obtained from the analytical wave functions that describe two noninteracting 
bosons, two noninteracting fermions, and the corresponding symmetrized wave function. 
We show that some of the properties of the interacting two-boson system resemble 
the noninteracting fermionic ones. This work is organized in the following way. 
First, in Section~\ref{sec2}, we present the considered Hamiltonian. In Section~\ref{sec3}, 
we present the analytic wave functions mentioned above. In Section~\ref{sec4}, we discuss 
how the ground state energy of the two-boson system changes as we vary the interaction. 
The corresponding effect on the density profiles and pair correlations is 
presented in Section~\ref{sec5}. In Section~\ref{sec6}, we briefly describe our 
numerical methods.  Finally, in Section~\ref{sec7}, we summarize the main 
conclusions of our work.

\section{The Hamiltonian}
\label{sec2}

The system of two interacting identical bosons in a two-dimensional isotropic 
harmonic trap is described by the Hamiltonian,

\begin{equation}
\label{Hamiltonian}
{\mathcal H}=\mathcal{H}_0
+ V(|\vec{x}_2-\vec{x}_1|)
\end{equation}
where the noninteracting part of the Hamiltonian reads

\begin{equation}
\label{Hamiltonian2}
\mathcal{H}_0=\sum_{i=1}^2 
\left ( -\frac{\hbar^2 }{2m}\nabla^2_i+\frac{1}{2}m\omega^2 |\vec{x}_i|^{\,2}\right ) \,,
\end{equation}
and the interaction is modeled by means of a Gaussian-shaped potential \cite{PRA96.043614,Doganov,Klaiman,Katsimiga,Bolsinger,Jeszenski},

\begin{equation}
\label{eqpot}
V(\vec{x}_1,\vec{x}_2)= \frac{g}{\pi s^2}e^{-\frac{|\vec{x}_2-\vec{x}_1|^2}{s^2}}\,.
\end{equation}
The interaction strength is controlled by $g$, while $s$ defines the interaction 
range. In this work, we always consider purely repulsive interactions, $g\geq 0$. A comparison between zero-range models and the Gaussian-shape potential has been studied in \cite{Doganov}. In the present work, we concentrate on the dependence on both  the strength $g$ and the range $s$ of the interaction. Notice, however, that we will not consider the regime of large scattering length, as in our case, with a purely repulsive interaction, the scattering length will be of the order of the range $s$ \cite{Doganov,Jeszenski}. The Hamiltonian can be split into the center-of-mass part and the relative part,

\begin{eqnarray}
{\mathcal H}_{\rm cm}&=&
-\frac{\hbar^2}{2\mathcal{M}}\nabla^2_{\vec{R}}+\frac{1}{2}\mathcal{M}\omega^2 R^{\,2} \,, \nonumber\\
{\mathcal H}_{\rm r}&=&-\frac{\hbar^2}{2\mu}\nabla^2_{\vec{r}}
+\frac{1}{2}\mu \omega^2 r^{\,2}+\frac{g}{\pi s^2}e^{-\frac{r^2}{s^2}}\,
\end{eqnarray}
where $\vec{R}\equiv\frac{1}{2}\left(\vec{x}_1+\vec{x}_2\right)$ is the center-of-mass 
coordinate, $\vec{r}\equiv\vec{x}_1-\vec{x}_2$ is the relative coordinate, 
$\mathcal{M}\equiv 2m$ is the total mass, and $\mu\equiv m/2$ is the reduced mass. 
Hereafter, we will use harmonic oscillator units, i.e. the length in units of 
$ \sqrt{\hbar/ (m\omega)}$ and the energy in units of $\hbar \omega$. The two 
pieces of the Hamiltonian in those units read

\begin{eqnarray}
{\mathcal H}_{\rm cm}&=&
-\frac{1}{4}\nabla^2_{\vec{R}}+R^{\,2}={\mathcal K}_{\rm cm}+{\mathcal V}^{cm}_{\rm ho} \nonumber\\
{\mathcal H}_{\rm r}&=&-\nabla^2_{\vec{r}}
+\frac{1}{4} r^{\,2}+\frac{g}{\pi s^2}e^{-\frac{r^2}{s^2}}
={\mathcal K}_{r}+{\mathcal V}^{r}_{\rm ho}+{\mathcal V}_{\rm int}
\label{eqcm2}
\end{eqnarray}
where we identify each part: the center-of-mass kinetic energy, 
${\mathcal K}_{cm}=-(1/4)\nabla^2_{\vec{R}}$, the relative kinetic energy, 
${\mathcal K}_{r}=-\nabla^2_{\vec{r}}$, the center-of-mass harmonic potential, 
${\mathcal V}^{cm}_{ho}=R^{\,2}$, the relative part of the harmonic potential, 
${\mathcal V}^{r}_{ho}=(1/4)r^{\,2}$, and the interaction term, 
${\mathcal V}_{int}=(g/(\pi s^2))\exp({-r^2/s^2})$, which has been properly 
transformed to the harmonic oscillator units.

\clearpage
\section{Analytic Wave Functions to Compare with the Ground State}
\label{sec3}

We are interested in understanding how the system changes its structure as we 
increase the interaction strength. In particular, we want to discern whether any 
fermionization takes place in the strong interaction limit, akin to the 1D case, 
despite the fact that the Bose--Fermi mapping theorem does not apply in 2D. To 
enlighten this discussion, we will compare the properties of our numerically 
obtained two-boson system with those of  (1) the wave function of the ground state 
of two bosons in the noninteracting limit, which will correspond to first 
order perturbation theory, (2) the wave function of the ground state 
of two fermions in the noninteracting limit, and (3) the wave function obtained by 
symmetrizing the previous one by taking its absolute value. We will use polar 
coordinates to express the wave functions.

\subsection{First-order Perturbation Theory}

The ground state of two noninteracting bosons under the Hamiltonian 
in~(\ref{Hamiltonian}) is the nondegenerate~state, 

\begin{equation}
\label{eqB}
\Psi_{B}(R,\varphi_R,r,\varphi_r)=\frac{1}{\pi}e^{-R^2-\frac{r^2}{4}},
\end{equation}
which is a state with zero angular momentum. Its energy is computed taking into account 
that it is an eigenfunction of the Hamiltonian in the noninteracting case, 
$\mathcal{H}_0\Psi_{B}=2\Psi_{B}$. The expectation value of the 
interaction term is

\begin{equation}
\langle \mathcal{V}_{\rm int} \rangle_{\Psi_{B}}=
\frac{g}{\pi s^2}4\pi^2\int_{0}^\infty R\, dR\, \int_{0}^\infty r\, dr\,
\frac{1}{\pi^2}e^{-2R^2-\frac{r^2}{2}}e^{-\frac{r^2}{s^2}}=\frac{g}{\pi (2+s^2)}.
\end{equation}
The total energy reads

\begin{equation}
\label{Benergy}
E_{B}=2+\frac{g}{\pi (s^2+2)},
\end{equation}
which is the first-order perturbation theory prediction for the energy of the 
system~\cite{PRA96.043614}. It is worth mentioning that the center-of-mass wave 
function contained in Equation~(\ref{eqB}) is an eigenfunction of $\mathcal{H}_{\rm cm}$ 
with an eigenvalue of $1$ energy unit. 

\subsection{The Non-Interacting Two-Fermion System}

For the noninteracting two-fermion system, the ground state would be 
two-fold degenerate, with a zero center-of-mass angular momentum, and the 
relative angular momentum equal to $1$ or $-1$:

\begin{equation}
\Psi^{\pm}_F(R,\varphi_R,r,\varphi_r)=\frac{1}{\pi\sqrt{2}}e^{-R^2-\frac{r^2}{4}}r e^{\pm i\varphi_r}\,.
\label{eq:fm}
\end{equation}
The previous states are also eigenstates of 
$\mathcal{H}_0$, $\mathcal{H}_0\Psi^{\pm}_{F}=3\Psi^{\pm}_{F}$. The expectation 
value of the interaction energy in this case is

\begin{equation}
\langle \mathcal{V}_{int}\rangle_{\Psi_{F}}=
\frac{g}{\pi s^2}4\pi^2\int_{0}^\infty R\, dR\, \int_{0}^\infty r\, dr\,
\frac{1}{2\pi^2}e^{-2R^2-\frac{r^2}{2}}r^2e^{-\frac{r^2}{s^2}}=g\frac{s^2}{\pi (2+s^2)^2},
\end{equation}
and the total energy is

\begin{equation}
\label{fenergy}
E_{F}=3+\frac{gs^2}{\pi (s^2+2)^2}\,.
\end{equation}
Let us note that this contribution vanishes, as it should, for zero-range interactions, $s=0$. 

\subsection{Bosonized Two-fermion System}

If we symmetrize the previous wave functions by taking their absolute value, 
$|\Psi^{\pm}_F|=\Psi_{|F|}$, we obtain bosonic wave functions which, as in 1D, do not allow
bosons to sit at the same position. Notice that both fermionic wave functions, (\ref{eq:fm}), are transformed into the same symmetric one:

\begin{equation}
\label{simmetrizedF}
\Psi_{|F|}(R,\varphi_R,r,\varphi_r)=\frac{1}{\pi\sqrt{2}}e^{-R^2-\frac{r^2}{4}}r,
\end{equation}
which has no angular dependence. The main effect of this symmetrization is that $\Psi_{|F|}$ 
is not an eigenfunction of $\mathcal{H}_0$, and the expectation value of the energy in the 
noninteracting case is $\langle \mathcal{H}_0\rangle_{\Psi_{|F|}}=5/2$. The interaction energy 
is the same as before, $\langle \mathcal{V}_{int}\rangle_{\Psi_{|F|}}=(g s^2)/(\pi (s^2+2)^2)$, 
and the expectation value of the total energy in this case is

\begin{equation}
\label{|f|energy}
E_{|F|}=\frac{5}{2}+\frac{gs^2}{\pi (s^2+2)^2} \,.
\end{equation}
The contribution of $\langle \mathcal{H}_0 \rangle$ to the previous expectation 
value, $\frac{5}{2}$, is not equally distributed between the kinetic and harmonic 
potential energy. Therefore, they do not fulfill the virial theorem, which is not necessary since $\Psi_{|F|}$ is not an eigenfunction of $\mathcal{H}_0$. The expected values for the center-of-mass kinetic and harmonic 
potential energies are 1/2 each one, in agreement with the fact that the center-of-mass 
part of $\Psi_{|F|}$ is an eigenfunction of ${\mathcal H}_{\rm cm}$ and it therefore
verifies the virial theorem, 
$\langle{\mathcal K}_{\rm cm}\rangle_{\Psi_{|F|}}=\langle{\mathcal V}^{\rm ho}_{\rm cm}\rangle_{\Psi_{|F|}}=1/2$.
For the kinetic energy of ${\mathcal H}_{\rm r}$ , we have to apply to the wave function the operator

\begin{equation}
{\mathcal K}_r=-\frac{\partial^2}{\partial r^2}
-\frac{1}{r}\frac{\partial}{\partial r}-\frac{1}{r^2}\frac{\partial^2}{\partial \varphi^2_r} \,,
\end{equation}
which results in

\begin{equation}
{\mathcal K}_r\Psi_{|F|}=-\frac{e^{-R^2-\frac{r^2}{4}}(4-8r^2+r^4)}{4\pi\sqrt{2}r} \,.
\end{equation}
The expectation value then is

\begin{equation}
\langle{\mathcal K}_r\rangle_{\Psi_{|F|}}=
-4\pi^2\int_{0}^\infty r\,dr\,\int_{0}^\infty R\,dR\, \frac{e^{-2R^2-\frac{r^2}{2}}(4-8r^2+r^4)}{8\pi^2}=1/2,
\end{equation}
and, for the harmonic potential relative energy,

\begin{equation}
\langle{\mathcal V}^r_{\rm ho}\rangle_{\Psi_{|F|}}=
4\pi^2\int_{0}^\infty r\,dr\,\int_{0}^\infty R\,dR\,\frac{1}{2\pi^2}e^{-2R^2-\frac{r^2}{2}}r^2\frac{r^2}{4}=1.
\end{equation}

To sum up, the total kinetic energy is $\langle {\mathcal K}\rangle_{\Psi_{|F|}}=1$, 
and the total harmonic potential energy is~$\langle {\mathcal V}_{\rm ho}\rangle_{\Psi_{|F|}}=3/2$. 

Let us stress that $\Psi_{|F|}$ can be shown to be the best variational wave function, 
in the strong interacting limit, of the form

\begin{equation}
\Phi_a(R,\varphi_R,r,\varphi_r)=\frac{1}{\pi \sqrt{2^a\Gamma(1+a)}}e^{-R^2-\frac{r^2}{4}}r^a.
\end{equation}
For $a=1$, we recover $\Psi_{|F|}=\Phi_1$. The set of functions are eigenfunctions of 
the center-of-mass part, ${\cal H}_{\rm cm}\Phi_a=1\Phi_a$. The relative kinetic energy 
is independent of the parameter $a$ and can be shown as $\langle {\mathcal K}_r\rangle_{\Phi_a}=1/2$, 
which coincides with $\langle {\mathcal K}_r\rangle_{\Psi_{|F|}}=\langle{\mathcal K}_r\rangle_{\Phi_a}$. 
The expected value of the relative harmonic oscillator potential energy is 
$\langle{\mathcal V}^r_{\rm ho}\rangle_{\Phi_a}=\frac{1+a}{2}$. The interaction energy reads

\begin{equation}
\langle {\mathcal V}_{\rm int}\rangle_{\Phi_a}=
g\frac{\left(1+\frac{2}{s^2}\right)^{-a}}{\pi(2+s^2)},
\end{equation}
which tends to zero when $s\rightarrow 0$ only if $a\geq 1$. The total variational 
energy depending on $a$ and $s$ is

\begin{equation}
E(a,s)_{\Phi_a}=1+ 1/2+\frac{1+a}{2}+g\frac{\left(1+\frac{2}{s^2}\right)^{-a}}{\pi(2+s^2)}.
\end{equation}

The energy minimum, in the short-range limit ($s\rightarrow 0$) and for strong 
interactions ($g\rightarrow \infty$), is reached when $a=1$, which is precisely the 
case $\Psi_{|F|}=\Phi_1$ and the energy is $E(a=1,s\rightarrow 0)_{\Phi_1}=E_{|F|}(s\rightarrow 0)=5/2$.


\section{The Interaction Effect in the Ground State Energy}
\label{sec4}
\subsection{The Energy Contributions}

The ground state energy of the interacting two-boson system is obtained numerically 
following the procedure described in~\citep{PRA96.043614}, an exact diagonalization 
of the Hamiltonian using the ARPACK implementation of the Lanczos algorithm. Here we 
supply additional information by splitting the total energy of the ground state 
into the kinetic, the harmonic potential, and the interaction contributions, in 
order to study, separately, the effect of increasing the interaction strength.

\begin{figure}[t]
\centering
\includegraphics[width=0.8\textwidth]{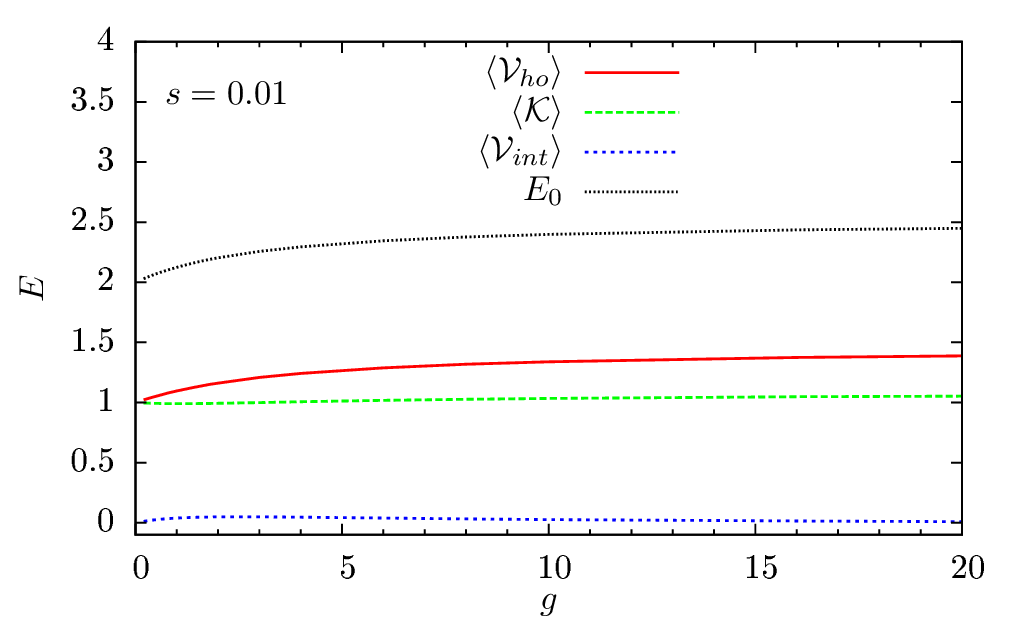}
\caption{Ground state energy of two interacting bosons in a two-dimensional harmonic trap, 
$E_0$, depending on the interaction strength, $g$, for a fixed and small 
interaction range, $s$. We present the harmonic potential part, $\langle{\mathcal V}_{\rm ho}\rangle$, 
the kinetic part, $\langle{\mathcal K}\rangle$, and the interaction part, $\langle{\mathcal V}_{\rm int}\rangle$. 
The energies were computed numerically using the first $M=200$ single-particle 
eigenstates of the harmonic oscillator (see Section~\ref{sec6} for details).}
\label{Fig:energies1}
\end{figure}

In Figure~\ref{Fig:energies1}, we see that, in the noninteracting limit, $g=0$, 
since the wave function of the ground state is the one given in~(\ref{eqB}), the 
total energy is $2$, arising from two equal contributions from the kinetic and 
potential energy. Increasing the interaction strength slightly affects  the 
interaction part of the energy. At the beginning, it starts to increase, but 
it remains mostly constant, decreases for larger interaction strengths, and 
becomes approximately $0$. This reflects that the particles avoid feeling the 
interaction by building dynamical quantum correlations in the wave function, 
as discussed in~\cite{PRA96.043614}. The most relevant contribution to the energy 
increase for a larger interaction strength comes from the harmonic potential part. 
The kinetic energy remains approximately constant and equal to $1$. Therefore, 
we can say that the two particles avoid feeling the interaction by separating 
one from the other. This reflects in an increase of the harmonic potential energy 
because they are further away from the center of the trap. This is consistent with 
the density profiles and pair correlation functions computed in~\citep{PRA96.043614} 
for different interaction strengths.

\subsection{Exploring the Strongly Interacting Limit for a Short-range Interaction}

In the strong interacting limit, in Figure~\ref{Fig:energies1}, we see that 
the total energy, the kinetic energy, and the potential energy tend to values 
corresponding to the symmetrized wave function of Equation~(\ref{simmetrizedF}). Clearly, 
the way the energy is distributed for the strongly interacting two-boson system in 
the zero-range limit does not coincide with the noninteracting two-fermion system 
but with the energy decomposition provided by $\Psi_{|F|}$. In addition, the value of the ground state energy seems to tend to 2.5 energy units, 
which is the expected value of the energy for the variational wave function~$\Psi_{|F|}$.

\begin{figure}[!bpth]
\centering
\includegraphics[width=0.8\textwidth]{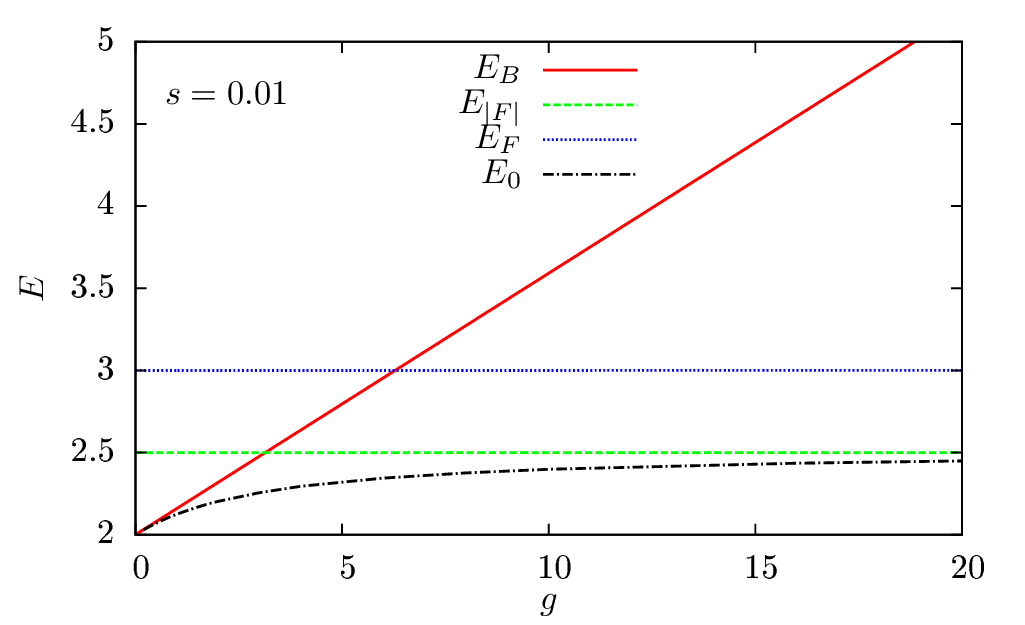}
\caption{Ground state energy of two interacting bosons in a harmonic trap, 
$E_0$, as in Figure~\ref{Fig:energies1}, depending on the interaction strength, 
$g$, and for a fixed range, $s$. The exact calculation is compared with the 
expectation value of the energy of the wave functions of two noninteracting 
bosons, $E_B$, two noninteracting fermions, $E_F$, and the corresponding 
symmetrized wavefunction, $E_{|F|}$, which are given, respectively, by Equations~(\ref{Benergy}), (\ref{fenergy}),~and (\ref{|f|energy}).}
\label{Fig:energiesFBF}
\end{figure}

In Figure~\ref{Fig:energiesFBF}, we see how the ground state energy computed numerically 
goes from $E_B$ for $g\simeq0$ to be very close to $E_{|F|}$ when $g$ becomes large for 
a small fixed range, $s=0.01$. $E_B$, $E_F$ and $E_{|F|}$ depend linearly on the 
interaction strength $g$ (see, respectively, Equations~(\ref{Benergy}),~(\ref{fenergy}),~(\ref{|f|energy})). 
However, the slope of the lines corresponding to $E_F$ and $E_{|F|}$ in Figure~\ref{Fig:energiesFBF} 
is positive but very small, due to the small value of $s$. 

\subsection{Exploring the Short-Range Limit for a Strong Interaction Strength}

\begin{figure}[!bhtp]
\centering
\includegraphics[width=0.8\textwidth]{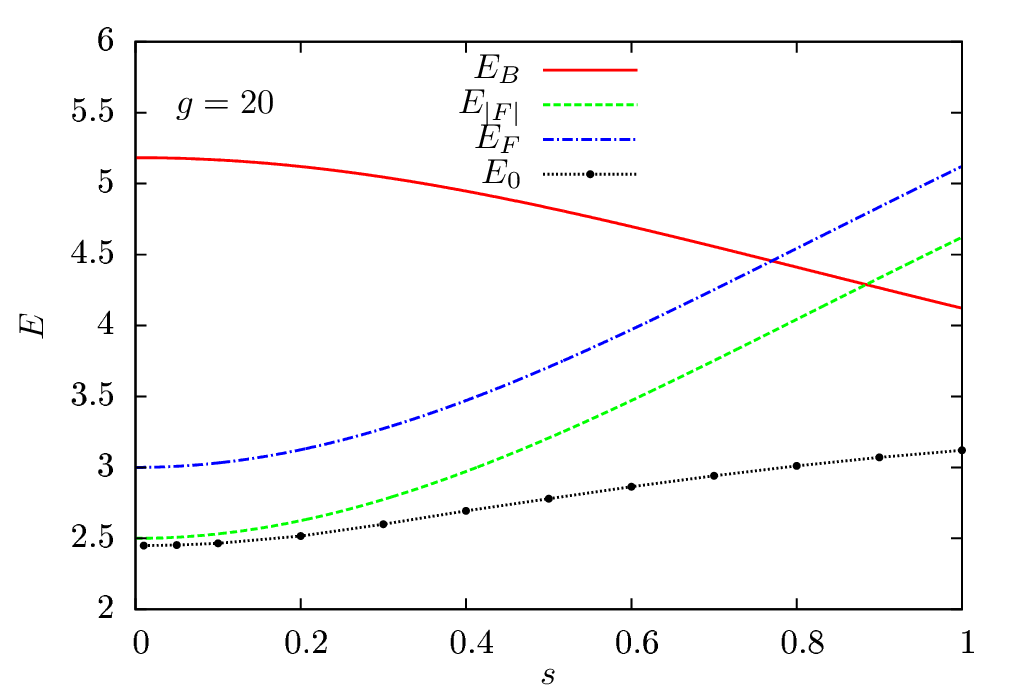}
\caption{Ground state energy of two interacting bosons in a harmonic trap, $E_0$, 
depending on the interaction range, $s$, and for a fixed interaction strength, $g$. 
The numerical result is compared with the expectation value of the energy of the 
wave functions of two noninteracting bosons, $E_B$, two noninteracting fermions, 
$E_F$, and its the symmetrized wavefunction, $E_{|F|}$, which are given, respectively, 
by Equations~(\ref{Benergy}),~(\ref{fenergy}),~and (\ref{|f|energy}).}
\label{Fig:senergies}
\end{figure}

The energy dependence on the range of the interaction is shown in Figure~\ref{Fig:senergies}, 
where we compare the ground state energy of the interacting two-boson system computed 
numerically with the expected values given by our analytic wave functions. For a 
fixed and strong interaction strength, $g=20$, reducing the range of the interaction 
results in a decrease of the ground state energy of the system. The same kind of behavior 
is observed for $\Psi_F$ and its symmetrized wave function, $\Psi_{|F|}$, since their 
dependence on the interaction is the same and the shift in energy between them is due 
to the noninteracting part of the Hamiltonian. In the short-range limit, their interaction 
energy tends to zero, and $E_0$ approaches $E_{|F|}$. Differently, the interaction energy 
of $\Psi_B$ does not vanish in the zero-range limit. In fact, the energy for this wave 
function increases for decreasing $s$.

\begin{figure}[!bhtp]
\centering
\includegraphics[width=0.8\textwidth]{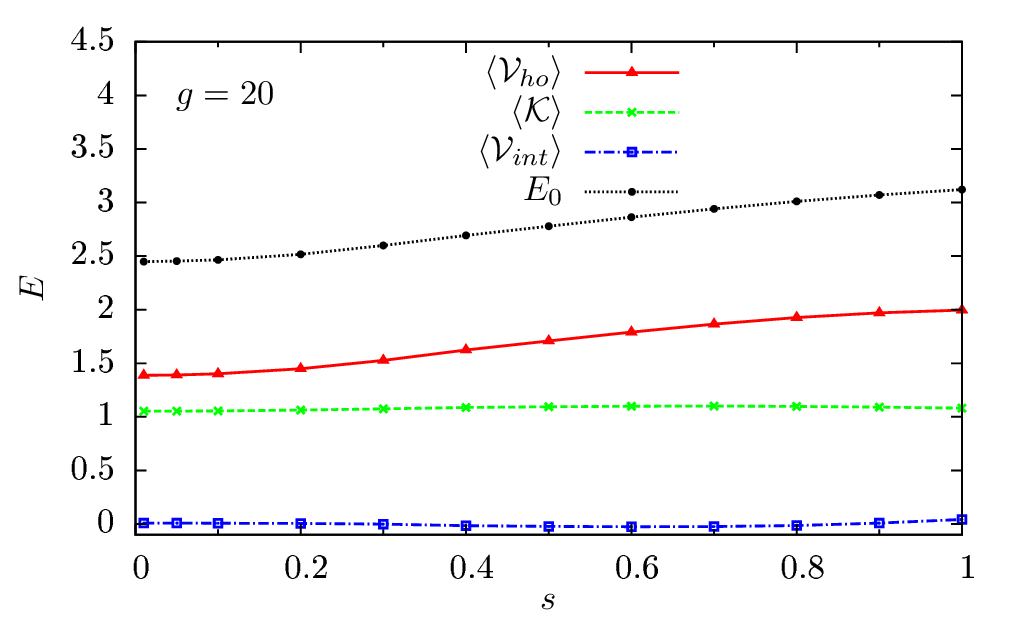}
\caption{Ground state energy of two interacting bosons in a harmonic trap, $E_0$, 
depending on the interaction range, $s$, for a fixed interaction strength, $g$. 
We also show the different contributions, $\langle{\mathcal V}_{\rm ho}\rangle$, 
$\langle{\mathcal K}\rangle$, and ${\langle\mathcal V}_{\rm int}\rangle$. The energies 
were computed numerically using the first $M=200$ single-particle eigenstates of 
the harmonic oscillator (see Section~\ref{sec6} for details).}
\label{Fig:energiessep}
\end{figure}

In Figure~\ref{Fig:energiessep}, we observe that not only is $E_0$ very close to $E_{|F|}$ 
when $s \rightarrow 0$, but the way the energy is distributed also coincides; i.e., 
the expected values of the kinetic energy, the interaction energy, and the harmonic 
potential energy for a strong interaction in the short-range limit for the ground state 
of the interacting two-boson system tend to values 
 computed for $\Psi_{|F|}$.

 In addition, we show that the most relevant contributions to the energy when the range 
$s$ varies comes from the harmonic potential energy, because the kinetic and interaction 
parts are mainly independent of $s$ and remain constant in Figure~\ref{Fig:energiessep}.

\section{The Density Profile and the Two-body Correlations}
\label{sec5}

In this section, we will compare the density profile and the pair correlation 
function of the ground state of the trapped two boson system, with and without 
interactions, with the ones provided by $\Psi_F$ and $\Psi_{|F|}$.

The density operator, acting on a system of $N$ particles, is defined as

\begin{equation}
\label{defdensityN}
\hat{\rho}(\vec{x})\equiv \frac{1}{N}\sum_{i=1}^N \delta({\vec{x}-\vec{x}_i}),
\end{equation}
and the pair correlation operator reads

\begin{equation}
\hat{\eta}(\vec{x},\vec{x}')\equiv 
\frac{1}{N(N-1)}\sum_{i=1}^N\sum_{j\neq i}^N \delta (\vec{x}-\vec{x}_i)\delta(\vec{x}'-\vec{x}_j)
\label{pairN}
\end{equation}
where both operators are normalized to unity. Since we consider the case of two 
identical particles, we compute the density profile and the pair correlation 
function for a given state $\Psi(\vec{x}_1,\vec{x}_2)$, respectively, as follows:
$\rho(\vec{x})=\int d\vec{x}_2\left | \Psi(\vec{x},\vec{x}_2) \right|^2$, and 
$\eta \left (\vec{x},\vec{x}'\right) = \left|\Psi \left (\vec{x},\vec{x}'\right) \right|^2$ . 

Using Cartesian coordinates, the wave functions of Section~\ref{sec3} read

\begin{eqnarray}
\Psi_{B} (\vec{x}_1,\vec{x}_2)&=&\frac{1}{\pi}e^{-\frac{1}{2}(x_1^2+y_1^2+x_2^2+y_2^2)}\,, \nonumber\\
\Psi_{F}^{\pm} (\vec{x}_1\vec{x}_2)&=&
\frac{1}{\pi \sqrt{2}}e^{-\frac{1}{2}(x_1^2+y_1^2+x_2^2+y_2^2)}\left((x_1-x_2)\pm i(y_1-y_2)\right)\,,\nonumber\\
\Psi_{|F|}(\vec{x}_1\vec{x}_2) &=&
\frac{1}{\pi \sqrt{2}}e^{-\frac{1}{2}(x_1^2+y_1^2+x_2^2+y_2^2)}\sqrt{(x_1-x_2)^2+(y_1-y_2)^2} \,.
\end{eqnarray}
The corresponding density profiles and pair correlations read

\begin{eqnarray}
\rho_B(\vec{x})                    &=&\frac{1}{\pi}e^{-(x^2+y^2)} \nonumber\\
\eta_B\left(\vec{x},\vec{x}'\right)&=&\frac{1}{\pi^2}e^{-(x^2+y^2+{x'}^2+{y'}^2)} \nonumber\\
\rho_F(\vec{x})                    &=&\rho_{|F|}(\vec{x})=\frac{1}{2\pi}e^{-(x^2+y^2)}(1+x^2+y^2) \nonumber\\
\eta_F(\vec{x},\vec{x}')&=&\eta_{|F|}(\vec{x},\vec{x}')=\frac{1}{2\pi^2}e^{-(x^2+y^2+{x'}^2+{y'}^2)} \left((x-x')^2+(y-y')^2\right) \,.
\end{eqnarray}
Note that both the fermionic and bosonized wave functions give the same density and pair correlation 
as $|\Psi_{|F|}(\vec{x}_1,\vec{x}_2)|^2=|\Psi_F^{\pm} (\vec{x}_1,\vec{x}_2)|^2$. 

From the pair correlation function and the density profile, we compute the probability 
of finding a particle at a distance $X\equiv \sqrt{x^2+y^2}$ once we have found the 
other at the origin, that is,

\begin{equation}
P(X;0)\equiv \frac{\eta(\vec{x},\vec{0})}{\rho(\vec{0})},
\end{equation}
which for the two previous discussed cases is, respectively, 
$P_B(X;0)=\eta_B(\vec{x},\vec{0})/\rho_B(\vec{0})=(1/\pi) e^{-X^2}$, and 
$P_F(X;0)=\eta_F(\vec{x},\vec{0})/\rho_F(\vec{0})=(1/\pi)e^{-X^2}X^2$.

\begin{figure}[!htbp]
\centering
\includegraphics[width=\textwidth]{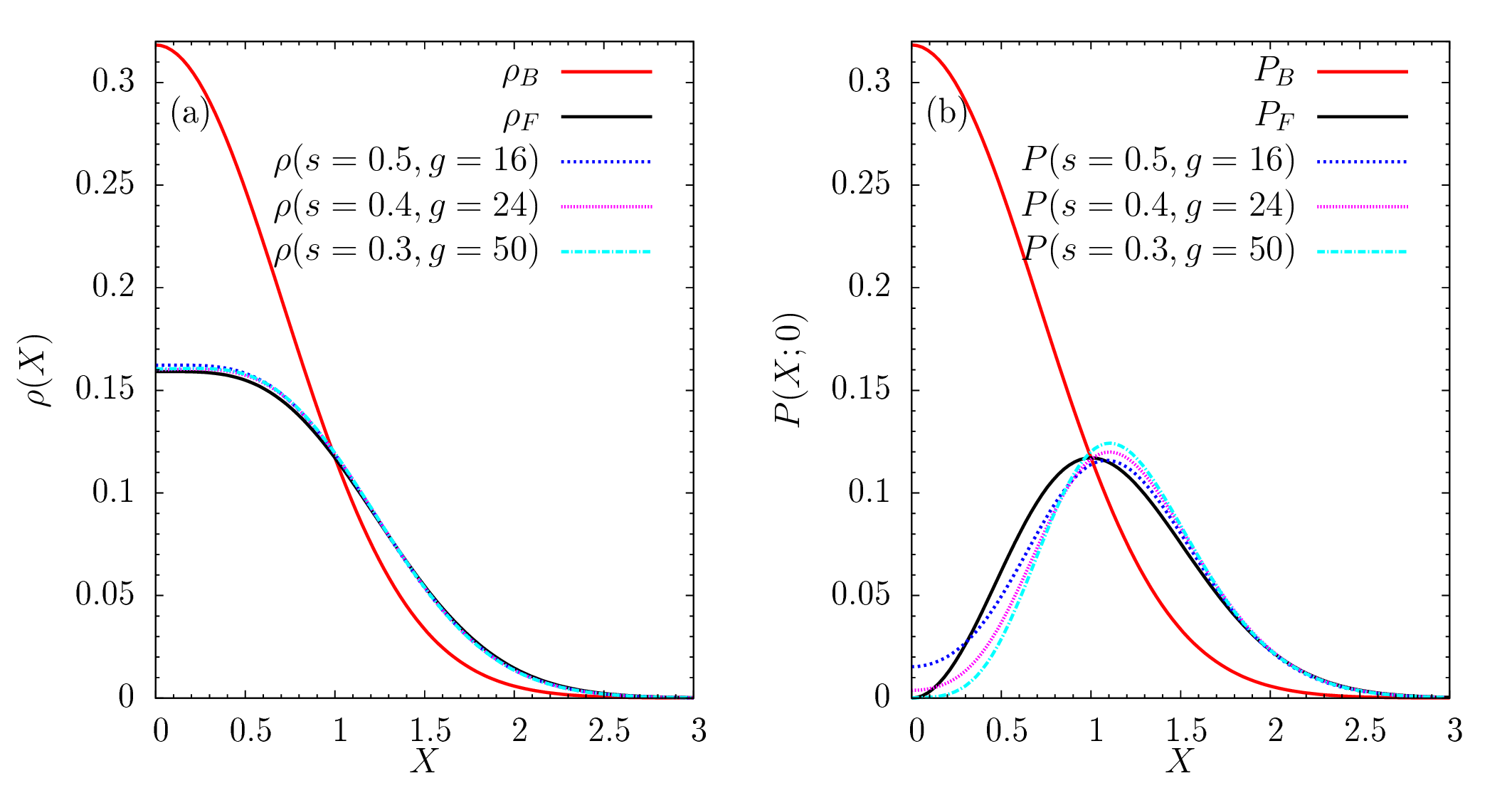}
\caption{(\textbf{a}) Density profiles of two noniteracting bosons, $\rho_B$, two 
noninteracting fermions, $\rho_F$, and three numerically computed profiles 
for different ranges and interaction strength for the interacting two-boson 
system. (\textbf{b}) Probability of finding a particle at a distance $X$ from the 
origin once a particle is found at $X=0$ in the same cases. For the numerical 
calculations, the number of single-particle eigenstates of the harmonic 
oscillator used was $M=80$ (see Section~\ref{sec6} for   details).}
\label{Fig:denspair}
\end{figure}

In Figure~\ref{Fig:denspair}a, we compare the density profiles obtained 
numerically for the ground state of trapped interacting two-boson system with 
the density profile corresponding to the noninteracting case and to the 
noninteracting two-fermion system. We show that, given an interaction range, 
there is an interaction strength such 
that the density profile of the interacting two-boson system is very well approximated 
by the two-fermion density profile. The smaller the interaction range, $s$, is, in which the density profiles coincide, the 
greater the interaction strength, $g$, will be. 
Therefore, in the short-range limit and for strong interactions, 
the density profile of two interacting bosons tends toward the noninteracting two-fermion profile.

In the case of the probability of finding a particle in space once we have 
found the other at the origin, we observe, in Figure~\ref{Fig:denspair}b, 
that the numerically computed ones for the interacting two-boson system 
resemble the corresponding   two noninteracting fermions. However, in this case, 
the maximum peak does not coincide, and it is closer to the center of the trap 
for two noninteracting fermions. The probability of finding the two bosons at 
the same position, i.e., at $X=0$, tends toward zero when the interaction range 
vanishes and the interaction strength increases, as expected.

\begin{figure}[t]
\centering
\includegraphics[width=\textwidth]{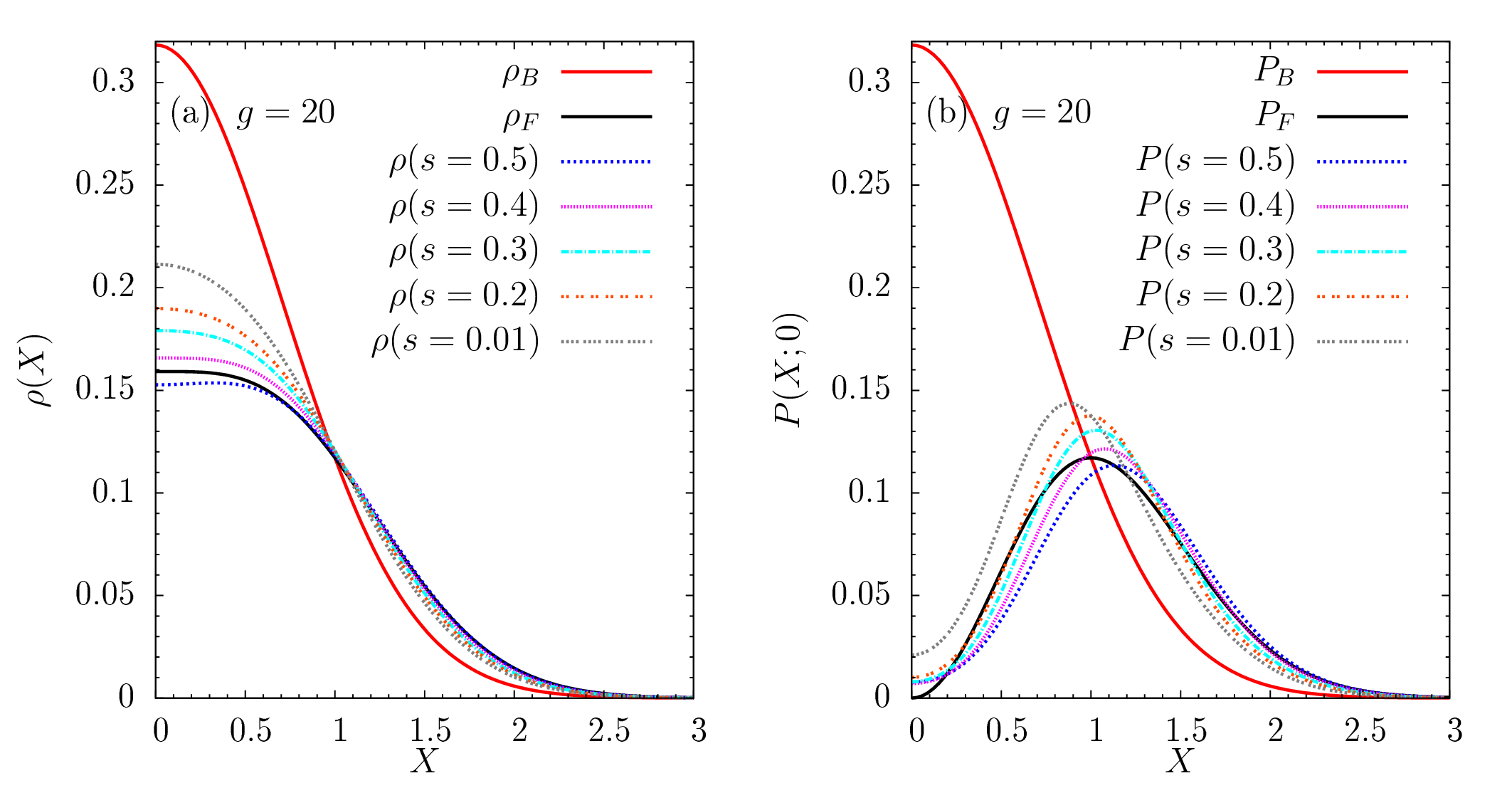}
\caption{(\textbf{a}) Density profiles of two noniteracting bosons, $\rho_B$, two noninteracting fermions, $\rho_F$, and four numerically computed profiles for different ranges fixing the interaction strength for the interacting two-boson system. (\textbf{b}) Probability of finding a particle at a distance $X$ from the origin once a particle is found at $X=0$ in the same cases. For the numerical calculations the number of single-particle eigenstates of the harmonic oscillator used was $M=80$ (see Section~\ref{sec6} for   details).}
\label{Fig:probX2}
\end{figure}

The effect of decreasing the interaction range for a fixed interaction strength 
is shown in Figure~\ref{Fig:probX2}. Regarding the density profile, in panel (a), 
the interaction strength is strong enough in all cases so that they are closer 
to $\rho_F$ than to $\rho_B$. The density peak increases 
 for a smaller range, 
although there is practically no effect in the density profile for $X>1$, and 
all of the numerical profiles fit well with $\rho_F$. In Figure~\ref{Fig:probX2}b, we observe that the effect of decreasing the range is such that the 
maximum in the probability approaches the center of the trap and the maximum 
value increases. However, in any case the probability vanishes at $X=0$, as 
it does for $\rho_F$, which would happen for greater $g$ values and in the 
short-range limit, as  shown in Figure~\ref{Fig:denspair}b.

\section{Numerical Method}
\label{sec6}
The numerical method was previously developed by the authors and is described in 
detail in~\cite{PRA96.043614}. First, we write the many-body Hamiltonian in 
a second quantized form. Then we define the set of single particle states, which 
are taken to be the lowest energy eigenstates of the single particle Hamiltonian. 
Once we fix the number of particles, we can define the corresponding Fock basis for 
the problem. 
The final results should be independent of the number of single particle states. In 
the numerical results reported in this paper, we have used up to 200 single particle states. 
The resulting Hamiltonian is then 
diagonalized using the ARPACK implementation of the Lanczos algorithm in order 
to compute the low-energy spectrum of the system and, in particular, the ground state.

\section{Conclusions}
\label{sec7}

In two dimensions, the ground state energy of two strongly interacting bosons 
in a harmonic trap for a short-range interaction is not equal to the noninteracting 
two-fermion system in the same potential, as it was in one dimension. However, the wave function resulting from 
symmetrizing the corresponding noninteracting fermionic ground state is found to be a very good variational trial wave function. It provides an upper-bound very close to the ground state energy obtained by exact diagonalization in the strong interacting and short-range limit. Even more, the distribution of the energy between 
the kinetic and harmonic potential parts in that limit also coincides. The increase of the energy with the interaction strength comes, mostly, from the harmonic potential 
contribution to the energy since the kinetic energy remains almost constant and the interaction 
energy approximately zero. The bosons avoid feeling the interaction by being 
more separate and, therefore, further from the center of the trap, which is 
also reflected in the density profiles obtained. The correlations induced by 
the interaction cause the density profile of the strongly interacting bosons to tend toward 
 the noninteracting two-fermion profile. For the pair correlation function, 
we have found qualitatively the same kind of behavior.

\vspace{6pt} 


\acknowledgments{The authors thank S. Pilati for his comments and discussions about the scattering length and the regularized pseudopotential in two dimensions. The authors acknowledge financial support of grants 2014SGR-401 from Generalitat de Catalunya and FIS2014-54672-P from the MINECO (Spain). P.M. was supported by an FI grant from Generalitat de Catalunya.}

\authorcontributions{Main ideas were conceived by Pere Mujal, Artur Polls, and Bruno Juli\'a-D\'iaz. Pere Mujal and Artur Polls performed all the calculations. Pere Mujal wrote a first version of the manuscript. Pere Mujal, Artur Polls and Bruno Juli\'a-D\'iaz discussed the results and wrote the final version of the manuscript.
}

\conflictsofinterest{The authors declare no conflict of interest.
} 

%
%
%

\reftitle{References}

%

%
%
%
\end{document}